# An Experiment and Detection Scheme for Cavity-based Light Cold Dark Matter Particle Searches

**Masroor H. S. Bukhari,[a, *] Zahoor H. Shah[b]**

[a] *Department of Physics, Faculty of Science, Jazan University, Gizan, Jazan 45142, Saudi Arabia.*
[b] *Faculty of Science, Malir University, Malir, Karachi, Pakistan*

## Abstract

A resonance detection scheme and some useful ideas for cavity-based searches of light cold dark matter particles (such as axions) are presented, as an effort to aid in the on-going endeavors in this direction as well as for future experiments, especially in possibly developing a table-top experiment. The scheme is based on our idea of a resonant detector, incorporating an integrated Tunnel Diode (TD) and a GaAs HEMT/HFET (High Electron Mobility Transistor/Heterogenous FET) transistor amplifier, weakly coupled to a cavity in a strong transverse magnetic field. The TD-amplifier combination is suggested as a sensitive and simple technique to facilitate resonance detection within the cavity while maintaining excellent noise performance, whereas our proposed Halbach magnet array could serve as a low-noise and permanent solution replacing the conventional electromagnets scheme. We present some preliminary test results which demonstrate resonance detection from simulated test signals in a small optimal axion mass range with superior Signal-to-Noise Ratios (SNR). Our suggested design also contains an overview of a simpler on-resonance *dc* signal read-out scheme replacing the complicated heterodyne read-out. We believe that all these factors and our propositions could possibly improve or at least simplify the resonance detection and read-out in cavity-based DM particle detection searches (and other spectroscopy applications) and reduce the complications (and associated costs), in addition to reducing the electromagnetic interference and background.

* Corresponding Author.
E-mail: mbukhari@jazanu.edu.sa

# 1. Introduction

Quantum Chromodynamics, or QCD [1], the theory of strong interactions, is known to conserve Charge conjugation and Parity (CP) operations, a fact which has long evaded cogent explanation. The QCD Lagrangian (similar to the Dirac equation) has inherently a conserved symmetry for CP, a fact known as the *"Strong CP problem"* [1, 2], for which a number of possible solutions or underlying reasons have been suggested. In this context, a suitable solution was proposed by Peccei and Quinn some time ago [3]. They suggested a global chiral U(1) symmetry, which is spontaneously broken to produce scalar particles (in a similar fashion to the production of Nambu-Goldstone bosons in SU(2) electroweak theory), a seemingly plausible solution to the strong CP problem. These solutions, the pseudo-NG bosons which couple to the gluon fields and break the symmetry, are known as *"Axions"* [4] and have been posited as possible candidates underlying the Dark Matter (DM) present throughout the universe. Depending on various models and theories, several kinds of axions have been proposed, which most notably include the so-called *"Hadronic axions"* or *KSVZ axions*, or the GUT (Grand Unified Theories) type or *DFSZ axions* [5]. The discussion on these is beyond the scope of this report, however, the axions the schemes in question attempt to explore are the former kind, henceforth referred to as KSVZ axions throughout this report. Since these axions belong to the non-luminous, low rest-mass range dark matter particles and are weakly interacting, these are known as "Cold Dark Matter (CDM)" axions.

In various theories for the existence of KSVZ axions possibly present in the DM halo of our galaxy, a strong possibility exists for these to be resonantly detected in a microwave cavity while subjected to a strong transverse magnetic field, under a variation of the *Inverse Primakoff effect* [6]. The Primakoff effect and its inverse, as illustrated in a Triangle diagram in Figure 1, have been understood for a long time as one of the several mechanisms producing photons from mesons and vice versa, in the form of a scattering process due to the charged particle's Coulomb field, ($\gamma$ + Ze ↔ Ze + A) [5, 6]. In such cases, if a large-scale macroscopic field, for instance, a magnetic field, is present, it has been suggested that the conversion can be seen as an axion-photon oscillation phenomenon [7] (similar to neutrino flavor oscillations), involving small momentum transfer and the interaction conserving coherence over a large distance [5].

Skivie [8] proposed in the early eighties that a similar production mechanism could exist for the production of axions in a strong magnetic field. Such axions, informally named as *"Skivie axions"*, are considered one of the strong possible candidates for light dark matter. Although there are many propositions and schemes reported in literature over last several decades, there are a few existing experiments in the process of development or being carried out [9, 10] which aim to explore these particles in cold resonant cavities while subjected to strong (on the order of several Tesla, typically 7.0-10.0T) magnetic fields. The aim of such searches is to create a resonant cavity environment with a high-quality cylindrical cavity (with support for the $TM_{010}$ modes) which is encapsulated in strong magnetic field transverse to the cavity axis and spectrally tuned to resonantly couple to any possible axions in situ, which could possibly result into production of resonant axion-photon events. While the cavity is tuned with a tuning rod across the range of frequencies corresponding to the axion mass range, the possible microwave photonic signals from any axion-photon conversions are picked up by an antenna and amplified by an amplification scheme comprising an RF SQUID (Superconducting Quantum Interference Device) and an HFET (Heterogeneous Field Effect Transistor) device [9], with the help of RF electronics and a slightly complex heterodyne read-out scheme [9]. In order to reach quantum-limited sensitivity and the desirable signal to noise ratio, the SQUID and HFET assembly are housed in a cryostat, operating at very low temperatures (from 4K all the way down to a few tens of milli-Kelvins). The signal is expected to lie within the microwave region (starting from GHz frequencies) corresponding to the posited axion mass.

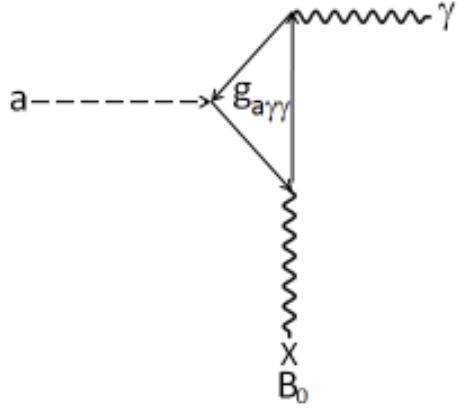

**Fig. 1.** Triangle diagram of the proposed axion-photon Conversion in a static magnetic field (B₀), under the so-called *Inverse Primakoff Effect*.

The Lagrangian for axion two-photon coupling may be expressed as [5]:

$$\mathcal{L}_{a\gamma\gamma} = -\frac{G_{a\gamma\gamma}}{4} F_{\mu\nu}\tilde{F}_{\mu\nu}\phi_a \qquad (1)$$

$$= G_{a\gamma\gamma}\boldsymbol{E}\cdot\boldsymbol{B}\phi_a \qquad (2)$$

Here $F_{\mu\nu}$ is the electromagnetic field strength tensor and $\tilde{F}_{\mu\nu}$ its dual, E and B are electromagnetic field intensities, $\phi_a$ is the axion field and $G_{a\gamma\gamma}$ is the axion two-photon coupling constant, expressed as:

$$G_{a\gamma\gamma} = \frac{\alpha}{2\pi f_a}\left(\frac{E}{N} - \frac{2(4+z)}{3(1+z)}\right) \qquad (3)$$

Where E and N are the electromagnetic and color anomalies of the axial current associated with the axion, $\alpha$ is the fine structure constant and $f_a$ is the axion decay constant (which can in turn be expressed in terms of a pion decay constant, $f_\pi$). The ratio in parantheses is the ratio of light quark masses and is approximately equal to 1.95, and the factor E/N is equal to 8/3 in the case of DFSZ and 0 in the case of KSVZ axions [11].

For a detailed discussion of the underlying theory, the reader is referred to [5] or any standard text on QCD.

The axion mass is calculated as [5];

$$m_a \approx 6\mu eV \frac{10^{12} GeV}{f_a} \qquad \text{\_\_\_\_\_\_\_\_\_\_\_} \qquad (4)$$

Ignoring the axion kinetic energy, the mass corresponds to a frequency of;

$$f_a = \frac{m_a c^2}{h} \qquad \text{\_\_\_\_\_\_\_\_\_\_\_} \qquad (5)$$

Thus, the detection of a signal in cavity with a frequency corresponding to the expected axion mass with a significant SNR may be taken as the signature of an axion. Although there are a number of challenges in these schemes, such as suppression of all possible background and maintaining the low-temperature environment as entailed by the experiment, one important challenge is related to the instrumentation to measure resonant signals corresponding to axion-photon events. One would like to have the most optimal detection scheme, whereby the required resonance signal is measured with the least possible noise. SQUID's are undoubtedly one of the most sensitive sensors to detect weak signals, to the lowest orders of power possible (~$10^{-18}$W), however they cannot facilitate the best possible detection of RF/*ac* resonance. A scheme has to be devised which may aid a SQUID in detection of weak resonance. Secondly, the instrumentation involved with RF signals measurement at high frequencies (such as various heterodyne schemes) is quite complicated and involved. Third, the superconducting electromagnets employed to provide the required magnetic field are expensive and noisy, requiring tedious operation and maintenance procedures. Therefore, a scheme which could possibly aid in these areas by new innovations or improvements on existing methods may be quite desirable in the on-going searches.

Working on these motivations, we propose a detection scheme by presenting some ideas such as a more sensitive resonance detection methodology, permanent magnetic field generation and a new simpler read-out scheme, which might aid in measuring on-resonance signals from possible axion-photon conversion events, while offering possible improvements in all these areas. A new and innovative detection element (the TD-FET duo) is designed and proposed to work in conjunction with a SQUID device to aid in more sensitive resonance detection by detecting voltage fluctuations in the TD-FET circuit as part of an LC tank; a simpler *dc* measurement method replaces the complicated high-frequency RF detection scheme; and we suggest a halbach permanent magnet array to offer an alternative to the electromagnet option (however, this is just a minor improvement, the scheme could still work with the conventional methods as employed in on-going searches).

To begin with, we suggest a hypothesis that axions could follow a mass distribution function and thus a given flux of axions could be populated by a distribution of axion masses. Since there is not much known about the axions and they remain a hypothetical particle, a particular distribution cannot be attributed to them. With this assumption, we intend to scan a small region of axion mass and concentrate on that, giving maximum effort to base our search in that region.

For a frequency ($f_a$) corresponding to 2.0GHz to 2.99GHz, we have an anticipated axion mass ($m_a$) range of approximately $8 \times 10^{-6}$ to $12.5 \times 10^{-6}$ eV. Using such models, we anticipate a projected halo axion density ($\rho_a$) of around 0.1-0.3GeV/cm$^3$. The ideal working temperature ($T_c$ or $T_{phys}$) for the detector and cavity has to be kept around 10-20mK whereas the ambient temperature for amplifiers is to be maintained at less than 4K with their noise temperatures not exceeding 2K. These constitute the basic parameters for the possible implementation of our simple (and yet phenomenological) model detection scheme to test its viability, as a proof of the principle, however these do not constitute as a complete set of parameters or mandatory ingredients of an actual dark matter experiment.

## 2. The Experiment

### 2.1 Overview

Working on the similar lines of conventional cavity-based light dark matter experiments, we propose here a modified experiment and the associated detection scheme involving a new detector design, with a hope that it might offer a valuable detection strategy for the postulated Skivie axions in the prescribed axion mass range. The salient features of this scheme are, at first, an enhanced resonance detection methodology, secondly, a simpler magnetic field solution using an array of permanent halbach magnets, and finally, a simpler *dc* read-out scheme. Figure 2 illustrates an overview schematic of our proposed experiment. The design of the detector is based on two elements, first, the use of a tunnel diode (TD) [12] installed at the cavity along with the SQUID serving as the main detector, and second, a cryogenic Ultra-Low Noise Amplifier (ULNA), a modified low-temperature and very-low noise form of a conventional Low-Noise Amplifier (LNA) [13], working in closed loop with the front-end detector. This ULNA is a marginal oscillator design [14] integrated with the TD detector, working as a Combo TD-ULNA detector-amplifier for enhanced, or possibly the best possible, resonance detection.

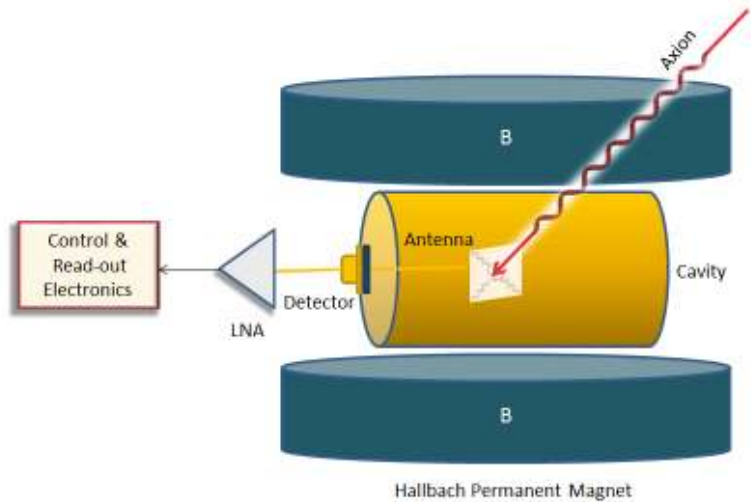

**Fig. 2.** Overview of the proposed experiment(and a cartoon of axion-photon conversion in the cavity) built around a resonant RF cavity and a magnet assembly. The detector is a Tunnel Diode and LNA is a Low-Noise Amplifier, whereas the magnet assembly is an array of Halbach permanent magnets.

As per our design, the TD becomes an integral part of the ULNA oscillator, which together with the cavity LC tank, creates the resonance in conjunction with the electromagnetic fields present in the cavity, thus in effect enhancing the power of any faint photons produced within as a result of possible axion conversions. By exploiting the negative conductance features of the integrated TD-HEMT oscillator, one can achieve a sensitive and simple resonance detection methodology. Since a tunnel diode is a very-low noise device, especially at around 10mK ambient temperature, the total amplifier noise is not more than any other HEMT based LNA scheme. The noise temperature can be reduced further by using a cascode configuration or by employing multiple amplifiers working in parallel to reduce noise statistically.

The RF amplification and read-out system are simplified in our design by incorporating a *dc* voltage monitoring and resonance detection (instead of an RF heterodyne scheme) to detect the on-resonance signals in a less involved manner. The room-temperature amplification, filtering and control schemes amplify the signal as well as provide with the required bias points for the TD and LNA, and present a conditioned signal in suitable form for read-out. We believe that these two elements working in tandem may offer a superior and more sensitive scheme for detecting resonant signals within the cavity, such as arising from the detection of Skivie axions (or any other particles which could be produced from an inverse Primakoff effect in a resonant cavity).

## 2.1 The Cavity

The central element of the experiment is the cavity, a high quality factor RF cavity made with either copper or a suitable superconducting element/alloy, such as Niobium or Niobium-Copper, with a cylindrical geometry to support the $TM_{010}$ modes (permitting parallel electric and magnetic field components). This mode facilitates highest electric field distribution in the cavity's center and helps obtain highest possible values of the interaction Lagrangian and maximum conversion power. Our proposed design is based on a 99.99% copper cylindrical cavity with special surface finish and electropolishing, as proposed elsewhere [15 and 16], to obtain a Q-value of at least $10^6$. However, realistically, we can achieve at this point a Q factor on the order of around $10^4$ -$10^5$ for cavities developed in our labs, which is quite good for reaching the sensitivity for detecting KSVZ axions as pointed elsewhere [17]. Moreover, the cavity volume has to be kept sufficiently large enough (in addition to maximizing the integration time) to permit highest conversion power [18]. Since our objective is to achieve a small-scale axion detection experiment (if not table-top at this stage), we propose a cavity with dimensions which could fit in an average sized room while making a small trade-off on conversion power. Our proposed cavity designs involve dimensions of 1 m length, inner diameter of 20cm and a wall thickness of 0.5mm, using 99% copper sheets, which would undergo a professional polishing process to increase their finesse and Q value. This would restrict the detection of axions to a smaller window of mass acceptance, but would be a good starting point to evaluate our proposed scheme.

## 2.2 The Magnet

In terms of the required axial magnetic field for the conversion cavity, we propose the use of an axial halbach geometry [19] permanent magnet array instead of an electromagnet in the hope to reduce the electromagnetic interference and complications (as well as running costs) in the experiment (However, the conventional superconducting electromagnets-based scheme could still be used.) The salient feature of such a magnet is that all the magnetic field is concentrated inside the bore of the magnet while there is zero field outside. This also helps eliminate the need of a bucking coil [9] in the cavity and other hardware to shield the SQUID's from magnetic fields. Such a magnet can be easily developed at any laboratory using a collection of rare earth (usually Neodymium-Iron-Boron alloy, NdFeB) magnets installed in a metallic frame following the halbach geometry patterns [19].

Halbach array magnets have been around since their introduction in particle accelerators for last one or two decades and have already made their way into Nuclear Magnetic Resonance (NMR) spectrometers and Magnetic Resonance Imaging (MRI) systems. Production of concentrated fields from 4.0T [20] to about 28.0T [21] have been made possible with such magnets.

Our proposed magnet array comprises ten magnet plates to make an axial magnet assembly as entailed for the experiment, with the cavity within the bore of the assembly. Each plate is made of eight Grade N52 rare earth (NdFeB) magnets [22] (dimensions 40x40x20 mm$^3$or 20x20x20mm$^3$), each with a magnetic field intensity of approximately 1.48T$\pm$0.1T or roughly 14,800G [23]. The magnet elements are fitted in a Halbach geometry in a robust toroidal steel reinforced aluminum fixture, and enclosed by a sturdy brass or copper ring which contains the powerful magnets (as per our design under development at the moment). This way a cylindrical magnet assembly is obtained which has placed within its bore the RF cavity for the experiment (similar to conventional axion experiments). The ratio of OD (Outer Diameter) to ID (Inner Diameter) is approximately 3 to 4. Figure 3 illustrates our preliminary design. By the virtue of the particular Halbach pattern geometry we choose, all the magnetic field is concentrated in the center with no field outside.This helps in creating a concentrated field inside and zero field outside, eliminating interference with external electronics. The aim is to obtain a 6.0T homogeneous magnetic field within the bore of the magnet, enclosing the cavity.

The magnetic field inside the halbach magnet bore is quite homogenous and can be calculated by using the relationship for this case (known as the "k=2" geometry);

$$H = M_r \ln \left( \frac{R_0}{R_i} \right) \hat{z} \ \_\_\_\_\_\_\_\_\_\_\_\_ \ (6)$$

Where M$_r$ is the ferromagnetic remanence, $R_i$ and $R_o$ are the inner and outer radii of the ring, respectively, and the field is in the z direction. It should be noted that if the ratio of the outer and inner radii exceeds the value of the natural constant $e$, the magnetic flux within the bore can reach its maximum and would be greater than the remanence of the material used in the construction of required magnet.

Multiple halbach rings like these can be built and would be stacked together in the z direction to obtain the cylinder for the experiment.

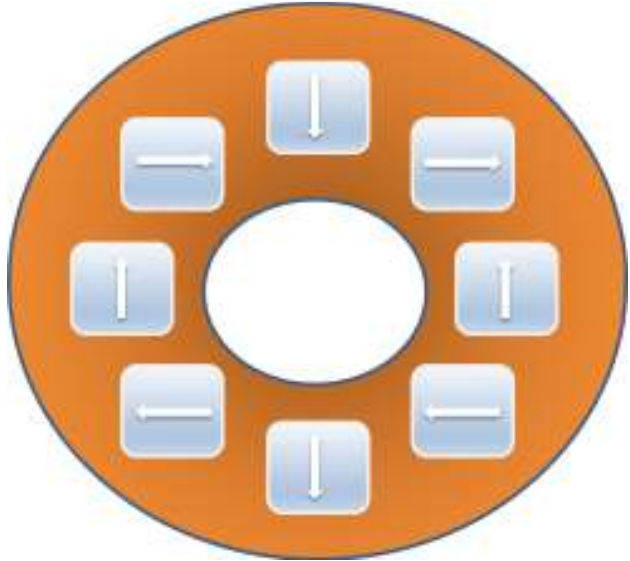

**Fig. 3.** A 1.48 ± 0.2T Halbach magnet array element made of eight N52 rare earth magnets, with cavity in the center, encapsulated between two robust steel or aluminum plates. A collection of these magnets aligned with each other along the z axis provide the required transverse magnetic field for the experiment.

### 2.3 Electronics

Figure 4 illustrates a block diagram for the electronics and read-out scheme associated with the experiment. The read-out electronics comprises a series of stages which aid in the data acquisition from the experiment and in providing necessary bias voltages and modulation signals.

As shown in the diagram, the faint resonant signal is obtained from the RF cavity (depicted here as a lumped LC element) with the help of a Superconducting QUantum Interference Device and a tunnel diode front-end detector and mixed with a 300KHz modulation signal from an ultra-low distortion Signal Generator in our laboratory (Stanford Research DS360). The signal is mixed with the help of a Varicap/Varactor Diode (VD) pair in parallel, which extends stability to the modulation signal, and transferred to the Low-Noise Amplifier (LNA/ULNA). There are two signals obtained from the output of LNA, a low-impedance *ac* RF signal and a dc voltage (the $V_d$, drain voltage of the LNA FET). The LNA is operated in its optimum conductance region under the control of an active bias circuitry (Amp1-Amp3). When the FET is maintained in its optimum region, a spike in the drain voltage corresponds to a resonant signal indicating the presence of resonance in the RF cavity. Since FET and the TD are sensitive to very small variations in the input current, even a weak signal owing to possible resonance in the LC tank circuit (the cavity and TD) can be picked up. This faint signal is further amplified by an RF amplifier, Amp3 (OP37) while removing the *dc* component from it and transferred to a 2MHz DSP Lock-In Amplifier (Stanford Research SR865; alternatively a 200MHZ SR844 could also be used for wider bandwidth), where the on-resonance signal is reconstructed. The LIA works in conjunction with the 300KHz modulation reference signal obtained from the signal generator to increase the sensitivity further. The FET's dc output also serves as the feed for the feedback amplifier to control the bias points for the FET and TD. Bias points and voltages are monitored in real-time with the help of a Measurement Computing MCC USB-1602HS-2AO Data Acquisition (DAQ), which is a 16-bit 2MSPS 2-BNC A/input channels quantization system, running with the National Instruments LabView software.

The tunnel diodes used in our prototypes are various Germanium models ranging from General Electric 1N3713 to 1N3717 [24] whereas the varactor diodes are the low-capacitance and hi-Q type Silicon Abrupt Junction Tuning Diodes (SMV1405 series, Skyworks inc.).

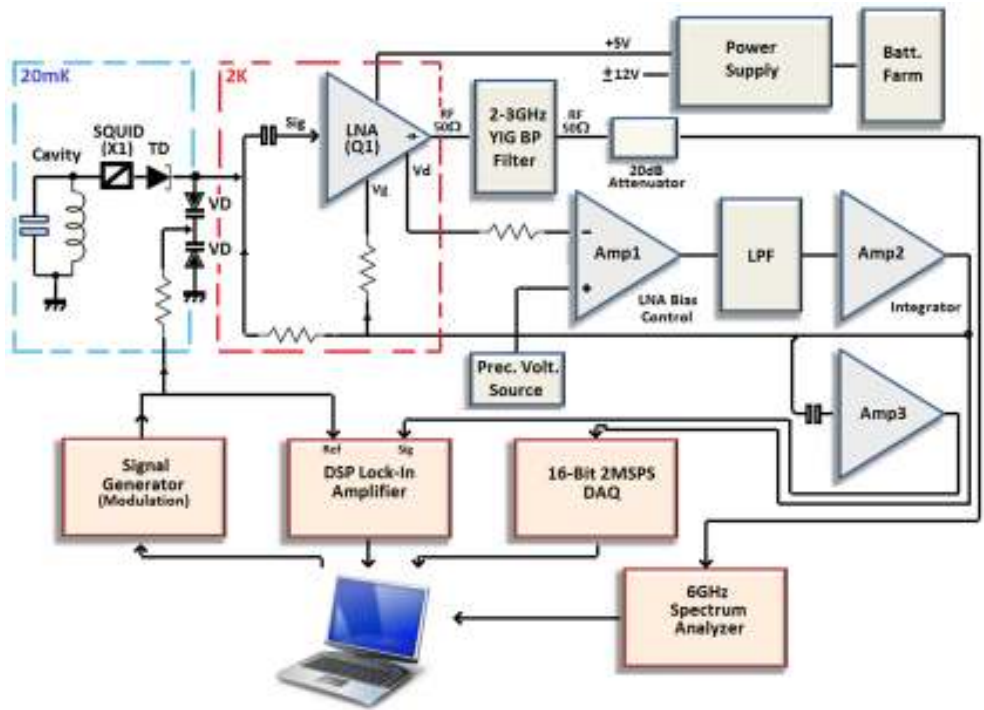

**Fig. 4.** Block diagram of electronics involved with the experiment including cavity (lumped LC model) and the Tunnel Diode (TD) detector. Note, VD denotes Varactor Diodes (Varicaps). The instrument is operated with a battery farm (+18V) to reduce interference and a regulated *dc* power supply produces +12V and +5V power rails required by the instrument.

The RF (*ac*) signal from the FET is filtered through a precision 2000-3000MHz band-pass YIG (Yttrium-Iron-Garnet) filter from Micro-Lambda Wireless, Fremont, CA (we use the model MLFS-1458, a dual-stage 2-4GHz YIG filter), and presented to a PC-based 6GHz spectrum analyzer (Aaronia GmBH Spectran HF6060v4) through a 20dB attenuator. This set of gear serves as our calibration analysis workstation as well as an online monitoring line for cavity real-time signal analysis. In parallel with our in-house developed LNA, we employed two commercial LNA's (1-4GHz band LNA models LNA3035 and LNA2030 by RFbay inc., Gaithesburg, MD) for testing and calibration of our RF system (these are commercial LNA's and hence there is neither a *dc* output available with those nor active biasing facility). The SQUID and TD setup must be ideally maintained at 20mK, whereas the LNA circuitry at 2K. The SQUID and LNA are prone to magnetic field interference (in contrast to the TD) and need to be kept in a non-magnetic environment, preferably using a zero-gauss chamber (or a bucking coil), which is one of the reasons we employ a Hallbach magnet array. As far as our post-amplifiers are concerned, we use a set of Thermo-Electric Cooling (TEC) plates to maintain these at about 240-250K for optimal low-noise operation. The experiment is maintained in an electromagnetically shielded solid copper-iron Faraday cage and covered with a special 70dB EM shield synthetic fiber canopy (*"70dB Shield Ultra"*, Aaronia GmBH).

## 3. Detector and Amplifier

### 3.1 Tunnel Diode Detector

A tunnel diode [12] is a highly doped semiconductor diode which has its *p* and *n* junctions heavily doped allowing a very large number of charge carriers to exist in the junctions. The structure allows a thin depletion layer formed in between, which enables quantum mechanical tunneling of electrons possible through this barrier. Even a small voltage applied to the diode results in a large current to flow, making it a unique and useful device. Some of its useful features include high-frequency (upto 3.6GHz our region of interest) and low-noise operation and ability to work at very low temperatures. A tunnel diode has very small inductance and capacitance and poses a differential negative conductance region between its minima and maxima. Once biased in this region (with the help of a constant *dc* bias voltage), the device acts as a sensitive oscillator and amplifier, offering detection and amplification of extremely weak *ac* signals, especially as part of an LC circuit, where it induces and detects oscillations at the resonant frequency. While used in a cavity and to take into account the effect of voltage drifts owing to varying temperatures etc., the negative conductance region has to be maintained by a feedback-based active biasing scheme. Thus, a tunnel diode seems to be an extremely sensitive and the most feasible device at a few GHz frequencies to detect RF resonance from a cavity or an LC circuit.

### 3.2 Ultra-Low Noise Amplifier (ULNA)

The on-resonance output signal from a tunnel diode is useful, however its magnitude is very weak and needs to be amplified at GHz frequencies. In such applications, an RF Low-Noise Amplifier (LNA) [13] is mandatory. In order to work in the application at hand, the required LNA has to offer very low noise (<1.0dB Noise Figure, NF, at 2-3GHz), highest possible gain (typically >15dB)  and high linearity, in addition to optimum stability, and moreover, it has to be able to be integrated with our TD circuit. All these specifications resulted into our design of an integrated TD-ULNA (Integrated Tunnel Diode-Ultra-Low Noise Amplifier) oscillator-detector-amplifier circuit, as modeled in a basic working schematic, Figure 5. This device is a modified form of a well-known FET marginal oscillator circuit (known at times as Colpitts Oscillator) [14] in which we have incorporated a TD with an FET (which is a High-Electron Mobility Transistor, HEMT [25], in this case). This is for the first time perhaps that a TD has been used in conjunction with an FET as a detector and oscillator, especially for resonance detection and amplification purposes. If properly designed and developed, this combination could serve as a powerful resonance detection scheme, especially for microwave frequencies.

We chose a GaAs Enhancement-mode pseudomorphic High Electron Mobility (EpHEMT) [26] LNA device from Avago (Agilent) for our ULNA design [27] (the ATF54143 [28]), in view of a number of factors, but most notably its pseudomorphic construction [45], excellent stability, high-linearity,reasonably low-noise and high-gain performance at few GHz (our region of interest) and GaAs construction which enables it to operate at cryogenic temperatures. It has been reported that pHEMT LNA's exhibit about one order of increase in electron mobility [29] than regular HEMTs, a factor which could be useful in our application. Moreover, this is a device which can be developed into a useful and precision LNA instrument at any small-scale laboratory like ours, not requiring any special facilities (Ideally, for an application of this magnitude, one could fabricate a specialized narrow-band MMIC [30] or HEMT device which has an integrated TD on its substrate, which would be a good idea for labs where such facilities exist.)

Similar to the tunnel diode, the HEMT entails a special bias region for its optimum operation, needing a passive or active bias scheme. Once operated in this region, the device could achieve sustained resonant oscillations in conjunction with the LC tank and TD. Thus, we had to come up with a design of an active biasing scheme for both the TD and HEMT where the individual bias points of both were maintained from one feedback control system. In our design [27], the device has been configured with passive RC bias networks to optimize its operation at about 2.75GHz center frequency which corresponds to the center of our aimed axion mass range. Working at this frequency it offers the required NF of <1.0dB and a gain of >15dB, with room for improvement. Both the design and development of this amplifier are not trivial and entail RF engineering and production of multiple prototypes before acquiring the required results. The details of the amplifier design and its construction (including the required special considerations for the HEMT) have been published in an earlier report [27].

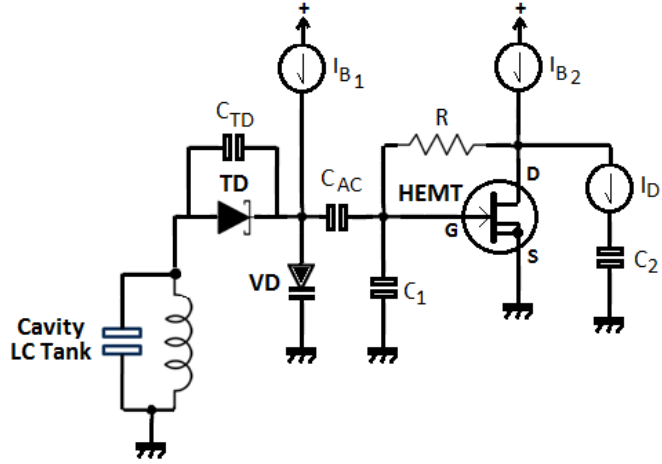

**Fig. 5.** A schematic of our modified TD-ULNA Oscillator-Detector-Amplifier. TD stands for Tunnel Diode, VD for Varactor Diode (Varicap) and HEMT for High-Electron Mobility Transistor. $I_{B1}$ and $I_{B2}$ refer to the bias currents for the TD and HEMT, respectively, whereas $I_D$ refers to the HEMT's drain current. R and $C_{TD}$ are the intrinsic gate-drain resistance and capacitance of the HEMT and TD, respectively.

Quantitative analysis and modeling of HEMT amplifiers as well as for a TD oscillator, especially for high-frequency operation with dc bias voltages in their optimal negative conductance regions have been discussed in detail elsewhere [46, 47]. Here, we limit our discussion to the detector-amplifier noise performance.

Following the definition of noise floor at room temperature,

$$Noise\ Floor\ (dBm) = TNF + NF + log_{10}(BW) \underline{\hspace{3cm}} \quad (7)$$

where TNF stands for Thermal Noise Floor (which is -174dBm), NF stands for the noise figure of LNA and BW the bandwidth,one can calculate the noise floor for the detector-amplifier.

At room temperature, the detector-amplifier has a noise voltage beginning with approximately 200nV, which translates to a noise floor on the order of $10^{-15}$W (or approximately -121dBm), at 100KHz bandwidth, whereas at low temperatures (0.025-0.1K) it is expected to reach its lowest sensitivity limit, approximately $10^{-18}$W (on the order of -151dBm), roughly the Minimum Detectable Signal (MDS) for this instrument in its present form. Similarly, while working at room temperature and at the center frequency of 2.75GHz, the amplifier offers a noise temperature of approximately 59K (corresponding to about 0.75dB NF), as expected. But it is brought down to approximately 2-2.5K at an ambient temperature of 1-2K, yielding an overall total system temperature of about 4-5K, which is a desirable noise temperature for cavity experiments.

Here an important aspect of the noise problem is the statistical reduction of noise. Over statistically large periods of sampling (owing to the Central limit theorem) an instrument with higher noise temperature could effectively be used to measure lower intensity signals. As argued in detail and shown heuristically by Partridge and Rohlfs [31 and 32], in the limit of large integration times, using an instrument with system noise temperature of 10K or more, one can venture down to a level of measuring faint signals like the Cosmic Microwave Background Radiation (CBR) which has an average temperature of 3K. As shown by [Partidge and Rohlf], the actual Minimum Detectable Signal (MDS) of a system (which is equivalent to RMS noise of a receiver) can be calculated by a form of the Radiometer equation [31]:

$$T_{rms} = T_{sys} \left[ \frac{1}{\Delta v \Delta t} + \left( \frac{\Delta G}{G} \right)^2 \right]^{1/2} \underline{\hspace{3cm}} (8)$$

Where $\Delta v$ is the total bandwidth of the system and $\Delta t$ measurement time.

If one attempts to measure a weak signal from an antenna using a radio, and the $\Delta G/G$ factor is kept very small tending to around $10^{-5}$, this translates to an antenna temperature as a function of simply the bandwidth and measurement time for a given signal $T_{sys}$.

$$(T_A)_{rms} \sim T_{sys} \left[ \frac{1}{\Delta v \Delta t} \right]^{1/2} \underline{\hspace{3cm}} (9)$$

Thus, for instance, one can even measure a faint signal of 1mK temperature from an antenna using an amplifier with a noise temperature of 10K, if one can sample long enough for a time duration of 1s [31].

In short, the TD and HEMT as part of a cavity act as a very sensitive cavity oscillations and resonance monitoring combination, which if developed properly, may offer an extremely valuable and low-noise resonance detection system.

We present our preliminary test results of the amplifier-detector system in the next section in the context of noise performance.

For further noise reduction, there are amplifiers available in the industry which may offer better noise performance than this particular design and could be customized and used in the conjunction with the detector for the problem at hand. A number of cryogenic LNA MMIC (Monolithic Microwave Integrated Circuit) designs using InP and other technologies for operation up to the W band [33] have been developed by the U.S. National Radio Astronomy Observatory (NRAO) and elsewhere. Such a device could be customized and employed in this application for better noise performance.

As suggested earlier, sensitivity of the experiment and its noise performance could be improved with the help of a SQUID amplifier. Design of such a device with potential for axion detection has been reported elsewhere [34], where a gain of 6-12dB and a noise temperature of 1-2K was reported at 2.2-4.0GHz while operated in a physical temperature of 4.2K, going down to about 50mK in a $T_{phys}$ of 20mK, but at a lower frequency of about 0.5GHz [35]. In recent times, there have been modified designs of SQUID RF amplifiers [36] with improved sensitivity and noise performance reaching the standard quantum limit and being capable of offering a quantum non-demolition interaction.

### 3.3 Post-Amplifier Electronics

The design of a post- amplifier is a modified form of an earlier design [27], having three stages; a buffer and isolator, an inverting feedback amplifier and an integrator and *ac* amplifier. These stages have been devised to monitor the bias point voltages at the input of HEMT and the tunnel diode, and based upon them, are used to maintain the bias voltages for the tunnel diode and HEMT to remain in the optimal region of their operation. Their other functions include the extract of the *dc* signal out of the ULNA amplifier for read-out and calibration, and to finally produce an *ac* signal for the Lock-in amplifier. Thesdesign of this section is on the lines of generic operational amplifier-based closed-loop feedback control systems for precision applications [37] (including those for integrating amplifiers as used in low-level signal measurements), for instance, as found in [38].

As illustrated in the Figure 4, an active feedback control circuitry (Amp1, Low-Passs Filter (LPF) and Amp2, made around two OP27 op-amplifiers, Analog Devices) amplifies and filters the obtained signal from the LNA, and based on that, maintains constant and balanced dc bias points for the TD ($V_{TD}$) and LNA (the $V_g$, gate voltage) to enable both of them to work in their negative conductance and optimum regions. A third amplifier, Amp3, an RF amplifier, made around a precision op-amplifier (Analog Devices OP37), recovers the *ac* signal and amplifies it for presenting to the Lock-in amplifier for final analysis and detection.

### 3.4 Test and Calibration

Test of the detection scheme by emulating expected axion events and measuring them, and before doing so, calibration of the setup are extremely important elements of the scheme.

For calibration of the involved detectors and amplifiers, we employ separate *ac* and *dc* calibration regimes. One of these schemes is also employed to simulate axion signal events in the cavity by generating and injecting (false) axion RF fields within the cavity  and attempting to measure the signals corresponding to it with our instrumentation.

Figure 6 illustrates various stages of our calibration scheme.

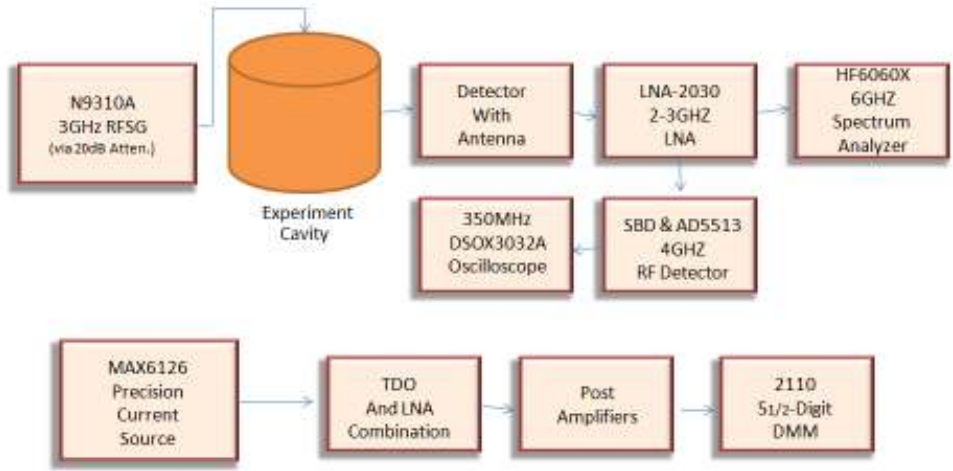

**Fig. 6.** Our adopted calibration scheme for the experiments.

For *ac* calibration and axion event emulation/simulation, an Agilent/Keysight N9310A 3GHz RF Signal Generator is used to generate calibration signals for the detector as well as to produce the test signal emulating an axion event in the cavity. This is used in conjunction with a standard RFBay LNA2030 LNA and an Aaronia HF4040 6GHz Spectrum Analyzer, to detect the signal as measured by the detector (using the RF output of the LNA). For *dc* calibration, especially for the resonance measurement, a test signal is weakly coupled to the cavity with the help of the RF Source (N9310A) and an antenna and the on-resonance *dc* voltages from the output of LNA are measured with the help of a standard RFD-5513 4MHz RF detector and an Agilent DSOX024A 200MHz 4-channel Digital Storage Oscilloscope for signal voltage measurement.

For calibration of power supplies, as well as for calibrating and monitoring the operating points of the tunnel diode and LNA, a precision voltage source made around MAX6126 and a Digital Multimeter (Keithley 2112) were deemed as the best method. The same oscilloscope (DSO) is utilized to analyze the power rail waveforms, at least once before every experiment, to check for any spurious signals or fluctuations, which is an important concern. Since we use battery power pack (with new batteries for actual experiments) for obtaining power for our setup, we have not observed any significant power rail fluctuations.

### 3.5 Detector-Amplifier Test Results

Although the scheme has not been implemented in an actual axion detection experiment so far, we present, at this stage, the test results of the detector-amplifier using high-frequency signals 'injected' into a model cavity to evaluate the scheme's suitability in resonance detection.

An axion 'event' of mass range 10-11μeV was simulated with the help of a 2.4-2.5GHz signal and weakly coupled to the cavity and it was attempted to detect resonance produced as a result of this signal, not by RF measurement, but by measuring the voltage changes owing to resonance with the help of our proposed detector-amplifier scheme. At this stage, we have not employed a SQUID device to ascertain the resonance detection capabilities of the detector.

A low-strength RF test signal from a source (an Agilent N9310) was weakly coupled to our model cavity, while the whole experiment (comprising the detector, ULNA and post-amplifiers, power supply and battery farm, excluding the measurement equipment and computers) was held in a high-grade pure copper Faraday cage and covered with an RF shield canopy. We carried out multiple experiments with no source signal to assure there was not significant noise or false signal detection. In all such experiment runs without an axion simulating signal coupled to the cavity there was no activity recorded at all except observing low-magnitude noise. Our noise suppression measures seemed to work effectively in eliminating most of the systemic noise. This was followed by the source signal weakly coupled to the cavity in identical conditions and a large number of experiments were carried out to detect and analyze what the scheme measured. Measurements were performed with two alternative instruments, a Digital Storage Oscilloscope (DSO) and a PC-based Data Acquisition (DAQ) scheme. We were able to successfully and consistently measure weak on-resonance signals using both schemes whenever a signal was sensed in the cavity from the source.

Figure 7a presents a plot of measured raw voltage for a test signal of 2.4GHz weakly coupled to a test cavity, as measured with the help of the Oscilloscope, whereas Figure 7b presents the conditioned voltages as measured with the DAQ setup. The data for these plots was obtained with identical experiment conditions and using a similar source signal, separated temporally by a few minutes. The plots reveal measured Signal-to-Noise levels of approximately 3 to around 4.5 at room temperature, which seem to be encouraging SNR levels. In the actual experiment, with the low-temperature operation, this ratio is expected to improve slightly, however, in view of the expected weak levels of the possible axion-induced events, one could expect signal-to-noise levels anywhere from 2-4.

At present, the sensitivity of the detector-amplifier, in terms of the axion mass it can detect, is ranging from 5 to 13μeV and centered at around 11μeV (without making any changes in the cavity), whereas its sensitivity in terms of the weakest signal it can measure is approximately -100 dBm ($\pm$10dBm) at the frequency of 2GHz (corresponding to an axion mass of around 13μeV) with a significant SNR (equal to or greater than 2), however the latter would improve greatly with the inclusion of a SQUID device in the detection stage, with sensitivity of the instrument reaching the levels of SQUID's minimum detectable signal.

The particular tunnel diode device, with which these results were obtained, was a GE 1N3713 device operated around an optimal bias point of 139mV$\pm$7mV, in conjunction with two parallel SMV1405 varactor diodes.

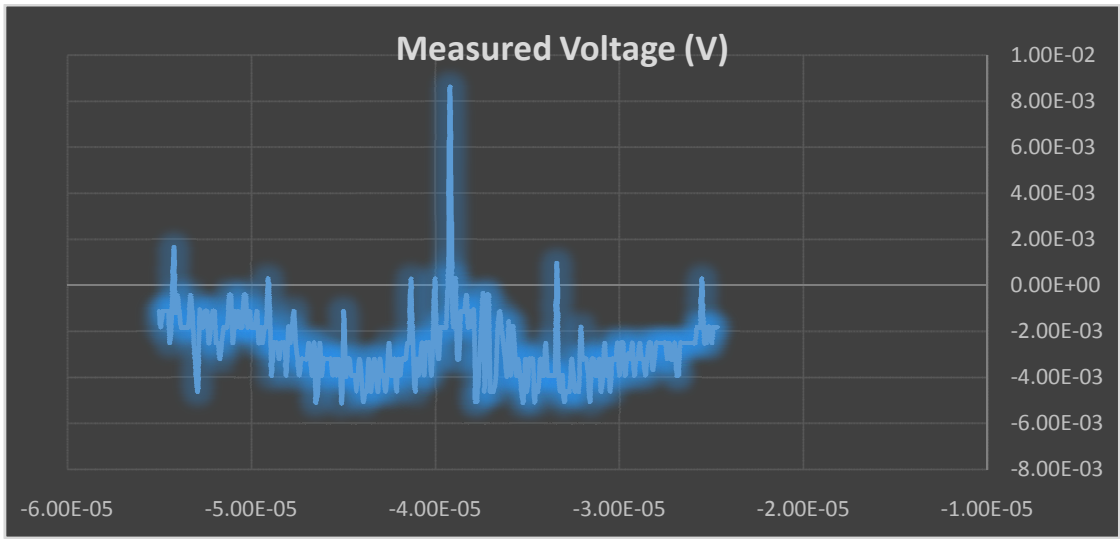

**Fig. 7a:** Raw dc voltages as a function of time measured with the setup corresponding to resonance from a 2.4GHz signal injected (weakly coupled) into the cavity, as detected by the TD and HEMT detector-amplifier at 300K and measured and displayed by our test Digital Storage Oscilloscope (DSO). The signal has roughly 8.5mVdc magnitude as expected and is a raw signal without any kind of signal conditioning (except the amplification and filtering in the on-board post-amplifier). One can observe a measured SNR of approximately 4.5 from this plot, which is encouraging at room temperature operation.

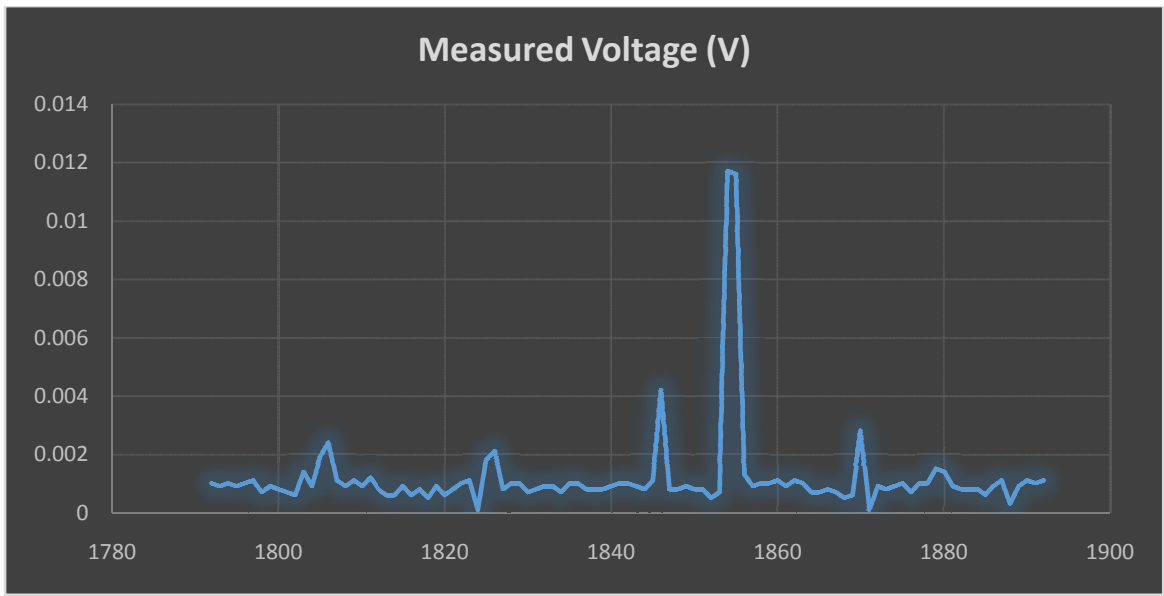

**Fig. 7b:** The measured voltages as obtained with the help of a Data Acquisition system (DAQ) scheme for identical experiments and an identical 2.45GHz source signal as shown in Figure 6a. A 12mV signal was measured by the DAQ scheme corresponding to resonance. Although the signal amplitude is quite higher, a slightly lower SNR (around 3) is obtained with the DAQ scheme as compared to the DSOX (greater than 4), which is a result of the production of noise from the signal conditioning stages in the DAQ.

## 4. Conclusions

We have presented a scheme and some ideas which could assist the cavity-based axion (or any other light cold dark matter particle) detection experiments underway in various parts of the world. The design and construction considerations are carefully devised to keep the noise levels and power consumption as well as experiment complications to a minimum, as discussed in this report.

It is encouraging to see that we have been able to measure with this instrumentation significant dc voltages corresponding to resonance signals falling within the cavity resonance frequencies, whereas presenting only very-low-strength flat noise for other frequencies, thus demonstrating in effect the potential of this scheme and the effectiveness of our measures for noise suppression, which in turn increase the sensitivity of the experiment. Excellent Signal to Noise Ratio's measured at room temperatures demonstrate the scheme's detection capability for detecting resonance with very little noise, such as from an axion-led event, if the particle exists and can be measured with a cavity-based inverse Primakoff effect scheme. In this context, to offer a comparison, we refer to the results reported elsewhere (Figure 4 in [9]) by a similar cavity-based axion search experiment, where a SNR of 0.5-1.5 was reported for much lower frequencies and at low temperatures. This is in itself an encouraging result that our designed resonance detection method could achieve twice the SNR at nearly quadruple frequencies.

In later stages, replacement of the GaAs pHEMT with an InP pHEMT [39] is planned to increase the amplifier's gain (it is reported that InP pseudomorphic designs have electron mobilities significantly higher than their GaAs counterparts [45]) and reduce the system noise temperatures further.

Some preliminary test results to demonstrate the proof-of-the-principle are presented here which seem quite satisfactory, although having room for improvement. The proposed detection method and amplification scheme may offer an efficient and simplified detection regime for the detection of on-resonance signals arising from the proposed Skivie axions, in contradistinction to the existing schemes. There are a number of other techniques for DM searches where this detection method could prove useful. This includes the Rydberg atoms-based cavity searches [17], where the instrument finds an immediate application.

We hope that the ideas presented here could be valuable in cavity-based dark matter search experiments, and possibly could develop into a table-top experiment for light DM particle searches in future.

This technique has potential in a number of areas other than cavity-based axion searches. First of all, since the Primakoff effect is not limited to the production of axions only it could be used in other branching modes too where photons could be produced from other particles. Secondly, the detector could be used in other areas of experimental physics, such as cavity-based cQED [40] and cirQED studies, analysis of ultra-cold systems [41] and various low-temperature spectroscopy techniques (such as NMR, ESR, EPR, etc.[42, 43, 44]), as well as in any other areas where weak signal resonance detection is desired.

This report is an endeavor to put forward some ideas, suggestions and the design of a technique for cold dark matter particle searches. It is hoped that the ideas presented here as well as the instrument developed in our labs could be a useful inspiration for the existing cavity-based axion and exotic particle search experiments around the world.

## Acknowledgments

Funding support from the King Abdul Aziz City of Science and Technology (KACST) Small Grant (Grant number 234-35) is acknowledged with gratitude which made this study possible. Authors would also like to thank the Deanship of Scientific Research (DSR), Jazan University for their help in setting up the laboratory where this research was conducted.